\documentclass[prl,twocolumn,floatfix,showkeys,showpacs ]{revtex4-1}
\UseRawInputEncoding
\usepackage{amsmath}
\usepackage{amssymb}
\usepackage{graphicx}
\usepackage{dcolumn}
\usepackage{bm}
\usepackage[colorlinks=true,linkcolor=blue,anchorcolor=blue, citecolor=cyan,urlcolor=cyan]{hyperref}
\usepackage[mathlines]{lineno}
\usepackage{ulem}
\usepackage{epstopdf}

\begin{document}
\title{External-field-induced transition from altermagnetic metal to fully-compensated ferrimagnetic metal in monolayer   $\mathrm{Cr_2O}$}
\author{San-Dong Guo$^{1}$$^{\textcolor[rgb]{0.00,0.00,1.00}{\dagger}}$}
\author{Qiqi Luo$^{1}$}
\author{Shi-Hao Zhang$^{2}$}
\author{Peng Jiang$^{3}$}
\affiliation{$^{1}$School of Electronic Engineering, Xi'an University of Posts and Telecommunications, Xi'an 710121, China}
\affiliation{$^{2}$School of Physics and Electronics, Hunan University, Changsha 410082, China}
\affiliation{$^3$Laboratory for Quantum Design of Functional Materials, and School of Physics and Electronic Engineering, Jiangsu Normal University, Xuzhou 221116, China}
\begin{abstract}
Altermagnets and fully-compensated ferrimagnets are two canonical classes of zero-net-moment magnets. An altermagnetic (AM) half-metal cannot exist due to its AM spin splitting, while a fully-compensated ferrimagnetic (FC-FIM) metal seems impossible to realize because both spin channels remain gapless.
Here,  we propose that an FC-FIM metal can be realized by breaking the rotational or mirror symmetry that links two spin-opposite magnetic atoms in an AM metal.
We further demonstrate that charge-carrier doping is fundamentally unable to generate a net magnetic moment in an altermagnet, whereas such a net moment can be readily induced in a fully-compensated ferrimagnet. We use the AM monolayer $\mathrm{Cr_2O}$ as a concrete example to validate our proposal. Either electric field or uniaxial strain can break the $S_{4z}$ symmetry of $\mathrm{Cr_2O}$, thereby inducing a transition from an AM metal to an FC-FIM metal.  Uniaxial strain plus carrier doping creates a net moment in an altermagnet, and  the so-called piezomagnetism  is essentially a strain-driven switch from altermagnetism to fully-compensated ferrimagnetism. By analogy, we advance the concept of electromagnetism: an electric field drives the transition from altermagnetism to fully-compensated ferrimagnetism, and subsequent charge-carrier doping stabilizes a net magnetization.
Our work provides a roadmap for further exploring the connection and distinction between altermagnet and fully-compensated ferrimagnet, and confirms the feasibility of FC-FIM metal.

\end{abstract}
\keywords{Fully-compensated ferrimagnetism, altermagnetism~~~~~~~~~~~~~~~~~~~~~~~~~~~~~$^{\textcolor[rgb]{0.00,0.00,1.00}{\dagger}}$sandongyuwang@163.com}

\maketitle
\textcolor[rgb]{0.00,0.00,1.00}{\textbf{Introduction.---}}
Emergent zero-net-magnetization magnets are attracting intense interest because they simultaneously deliver ultrahigh spintronic storage densities, robust immunity against external perturbations, and femtosecond-scale write speeds\cite{k1,k2}.
From a symmetry perspective \cite{k4,k5,f4,zg2}, collinear zero-net-magnetization magnets  encompass three principal classes: $PT$-antiferromagnet  (the joint symmetry ($PT$) of space inversion symmetry ($P$) and time-reversal symmetry ($T$)), altermagnet, and fully-compensated ferrimagnet.
Although $PT$-antiferromagnet exhibits vanishing net magnetization, their lack of spin splitting in momentum space imposes a fundamental constraint on practical deployment. Both altermagnet and fully-compensated ferrimagnet can  exhibit spin splitting without the assistance  of spin-orbital coupling (SOC) in momentum space, giving rise to a variety of intriguing physical phenomena, such  as the anomalous Hall and Nernst effects, non-relativistic spin-polarized currents, and the magneto-optical Kerr effects\cite{f4,zg1}.

 The two opposite-spin sublattices of  altermagnet are connected by  rotational/mirror ($C/M$) symmetry,  but the two spin sublattices  in fully-compensated ferrimagnet  are not connected by any symmetry\cite{k4,k5,f4,zg2,f5,f6,f7,f8}.  Experimentally, several bulk  altermagnetic (AM) materials have been confirmed  that exhibit momentum-dependent spin splitting\cite{k4,zg1,ex1,ex2,ex3,ex4}. Nevertheless, two-dimensional (2D) altermagnets remain a purely theoretical prediction\cite{zg2,k60,k601,k602, k6,k7,k7-1,k7-2,k7-3,k7-3-1,k7-3-10,k7-3-11,k7-3-12}. Fully-compensated ferrimagnetism has traditionally been pursued in bulk systems via chemical alloying\cite{f1,f2,f3}. In  2D materials, the fully-compensated ferrimagnetic (FC-FIM) magnet  can be electrostatically switched when the spin and layer are coupled with A-type antiferromagnetic (AFM)  ordering\cite{f4,zg2, gsd1,gsd2,gsd3,gsd4,qq5,qq6}, which has recently demonstrated in bilayer  $\mathrm{CrPS_4}$ under a perpendicular electric field\cite{nn}. Alternatively, the FC-FIM state can be accomplished without any lattice modification by re-engineering the spin ordering \cite{qq7},  or by stacking two ferromagnetic (FM) monolayers whose identical total moments exactly cancel across the heterojunction\cite{qq7-11}.

  As discussed in next  section, the spin-up and spin-down densities of states  in altermagnets are identical, which restricts AM phases to metal or semiconductor and effectively precludes the realization of an AM half-metal.
 In fully-compensated ferrimagnets, the spin-up and spin-down densities of states are generally unequal, thereby permitting the existence of metallic, semiconducting, and half-metallic phases. Nevertheless, the zero-net-moment of FC-FIM is not enforced by symmetry, and  in general, a gap must be maintained in at least one spin channel to guarantee compensation\cite{f4}. This constraint restricts FC-FIM phases to semiconductor or half-metal, seemingly ruling out an FC-FIM metal.
 However, achieving  an FC-FIM metal has been proposed by electrically closing the gap of a bilayer whose interlayer coupling is AFM and whose constituent monolayer is unipolar magnetic semiconductor (UMS)\cite{qq7-12}.
Here, we demonstrate that an FC-FIM metal can  also emerge  when the $C/M$ symmetry relating spin-opposite sublattices in an AM metal is  broken.
We validate the scenario by first-principles calculations performed on the AM monolayer $\mathrm{Cr_2O}$\cite{jp}.

\begin{figure}[t]
    \centering
    \includegraphics[width=0.46\textwidth]{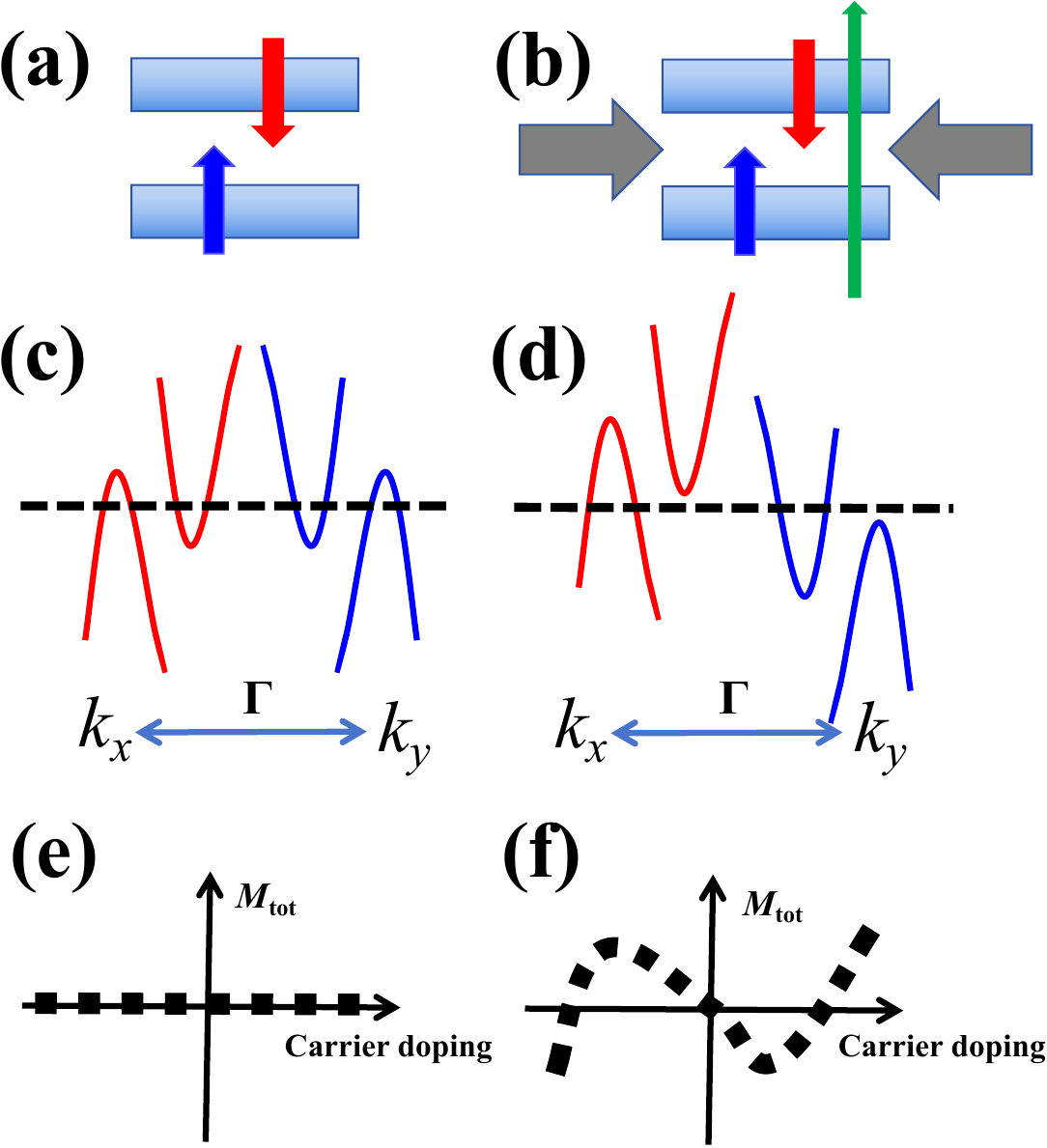}
    \caption{(Color online) (a): the A-type AFM AM tetragonal monolayer,  and the magnetic atoms with opposite spins are connected through [$C_2$$\parallel$$S_{4z}$] symmetry;  (b):  an out-of-plane electric field or in-plane uniaxial strain is applied in (a) to break  [$C_2$$\parallel$$S_{4z}$] symmetry; (c): the (a)  possesses   semi-metallic bands with AM spin splitting between $\Gamma$-$k_x$ and $\Gamma$-$k_y$ directions; (d): the (c) transitions into an FC-FIM metallic bands due to broken  [$C_2$$\parallel$$S_{4z}$] symmetry caused by an out-of-plane electric field or in-plane uniaxial strain; (e, f): by tuning the charge-carrier density with an electrostatic gate voltage, the AM metal (c) remains free of any net magnetic moment, whereas the FC-FIM metal (d) possesses a net moment that varies with carrier concentration. In (a, b), the blue, red, green, and horizontal gray arrows denote spin-up, spin-down, electric field, and uniaxial strain, respectively. In (c, d), the blue and red curves represent the spin-up and spin-down characteristic bands, respectively.}\label{a}
    \end{figure}

\textcolor[rgb]{0.00,0.00,1.00}{\textbf{Proposal.---}}
An FC-FIM metal can be obtained by breaking the $C/M$symmetry that connects spin-opposite magnetic atoms in an AM  metal.
Here we illustrate our proposal with an A -type AFM tetragonal AM  semimetal (\autoref{a} (a)). The magnetic atoms with opposite spins are connected through [$C_2$$\parallel$$S_{4z}$] symmetry  (with $S_{4z}$ being a roto-inversion operation),  showing   AM spin-splitting  between $\Gamma$-$k_x$ and $\Gamma$-$k_y$ directions (\autoref{a} (c)) with $d$-wave symmetry. When an out-of-plane electric field (Here, the external electric field can also be equivalently realized through the substrate or a ferroelectric heterojunction.) or in-plane uniaxial strain is applied (\autoref{a} (b)), the [$C_2$$\parallel$$S_{4z}$]  symmetry is broken, and  the system becomes an FC-FIM metal, showing global spin-splitting (\autoref{a} (d)) with $s$-wave symmetry.

For altermagnets, the spin-up and spin-down energy bands exhibit the following relationship:
\begin{equation}\label{d-1}
E_{\uparrow}(\vec{k})=[C_2\parallel O]E_{\uparrow}(\vec{k})=E_{\downarrow}(O\vec{k})
\end{equation}
where $O$ denotes $C/M$ symmetry. For the high-symmetry point $\Gamma$, we have $O\Gamma$=$\Gamma$, and then $E_{\uparrow}(\Gamma)$=$E_{\downarrow}(\Gamma)$. However, for a fully-compensated ferrimagnet, $E_{\uparrow}(\Gamma)$$\neq$$E_{\downarrow}(\Gamma)$ in general.

For altermagnets, according to \autoref{d-1}, we obtain:
\begin{equation}\label{d-2}
g_{\uparrow}(E)=\sum_{\vec{k}}\delta[E-E_{\uparrow}(\vec{k})]=\sum_{\vec{k}}\delta[E-E_{\downarrow}(O\vec{k})]
\end{equation}
When $\vec{k}$ spans the entire first Brillouin zone (BZ), $O\vec{k}$ also spans the entire BZ. Therefore, \autoref{d-2} can be rewritten as:
\begin{equation}\label{d-3}
\sum_{\vec{k}}\delta[E-E_{\downarrow}(O\vec{k})]=\sum_{\vec{k}^{\prime}}\delta[E-E_{\downarrow}(\vec{k}^{\prime})]=g_{\downarrow}(E)
\end{equation}
where $g_{\uparrow}(E)$ and  $g_{\downarrow}(E)$  denote spin-up and spin-down density of states.  According to \autoref{d-2} and \autoref{d-3}, $g_{\uparrow}(E)$ and  $g_{\downarrow}(E)$  are identical ($g_{\uparrow}(E)$=$g_{\downarrow}(E)$) in altermagnets, but they are generally not equal ($g_{\uparrow}(E)$$\neq$$g_{\downarrow}(E)$)  in fully-compensated ferrimagnets.

In general, the net magnetization  $M_{tot}$ in the magnet can be written as:
\begin{equation}\label{d-4}
M_{tot}=\mu_B\int_{-\infty}^{E_F}(g_{\uparrow}(E)-g_{\downarrow}(E))dE
\end{equation}
where $\mu_B$ is the Bohr magneton, $E_F$ is the Fermi level.
For both altermagnets and fully-compensated ferrimagnets,  $M_{tot}$ is equal to zero. Nevertheless, when shifting the Fermi level $E_F$ to vary the charge-carrier concentration, $M_{tot}$ remains exactly zero in altermagnets (\autoref{a} (e)) due to $g_{\uparrow}(E)$=$g_{\downarrow}(E)$ (In carrier doping, the magnetic ground state is assumed to remain unchanged.), whereas it is generally non-zero in fully-compensated ferrimagnets (\autoref{a} (f)) due to $g_{\uparrow}(E)$$\neq$$g_{\downarrow}(E)$.
Below, we use the AM  $\mathrm{Cr_2O}$ monolayer as an example to validate our proposal.

\begin{figure*}[t]
    \centering
    \includegraphics[width=0.92\textwidth]{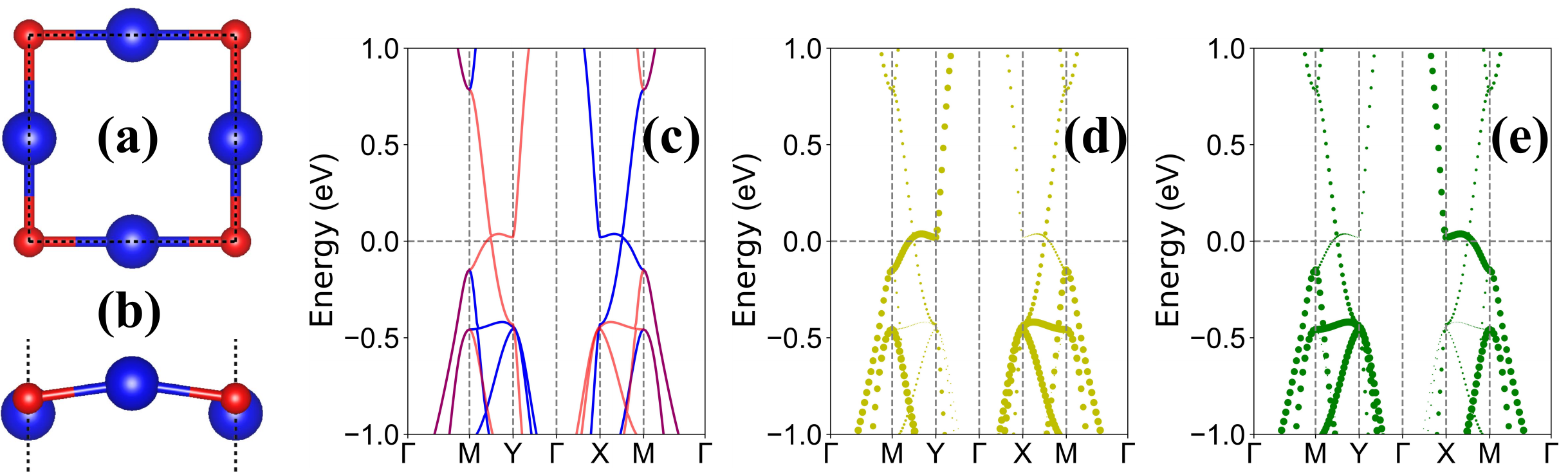}
    \caption{(Color online)For $\mathrm{Cr_2O}$, (a, b): the top and side views of the crystal structures; (c):  the spin-polarized band structures; (d, e): the upper- and lower-layer Cr projected band structures. In (a, b), the large blue and small red spheres represent Cr and O atoms, respectively. In (c), the blue and red curves denote the spin-up and spin-down characteristic bands, respectively, while the purple indicates the spin-degenerate bands. In (d, e), the size of the circles is proportional to the atomic weight.}\label{b}
\end{figure*}

\begin{figure*}[t]
    \centering
    \includegraphics[width=0.92\textwidth]{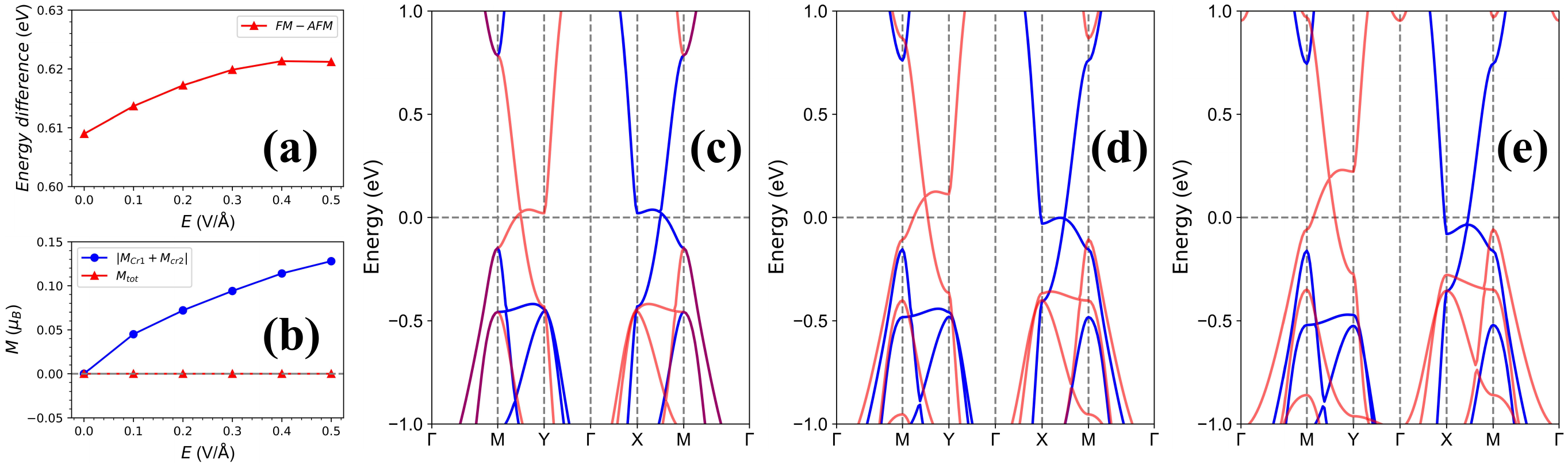}
    \caption{(Color online)For $\mathrm{Cr_2O}$, (a): the energy difference between FM and AFM orderings  as a function of electric field $E$; (b): the absolute value of the sum of the magnetic moments of the upper- and lower-layer Cr atoms ($|M_{Cr1}+M_{Cr2}|$), along with the total magnetic moment ($M_{tot}$), as a function of electric field $E$; (c, d, e): the spin-polarized band structures at  representative $E$=+0.00,  +0.20 and +0.40  $\mathrm{V/{\AA}}$.  In (c, d ,e), the blue and red curves denote the spin-up and spin-down characteristic bands, respectively, while the purple indicates the spin-degenerate bands.}\label{c}
\end{figure*}
\begin{figure}[t]
    \centering
    \includegraphics[width=0.46\textwidth]{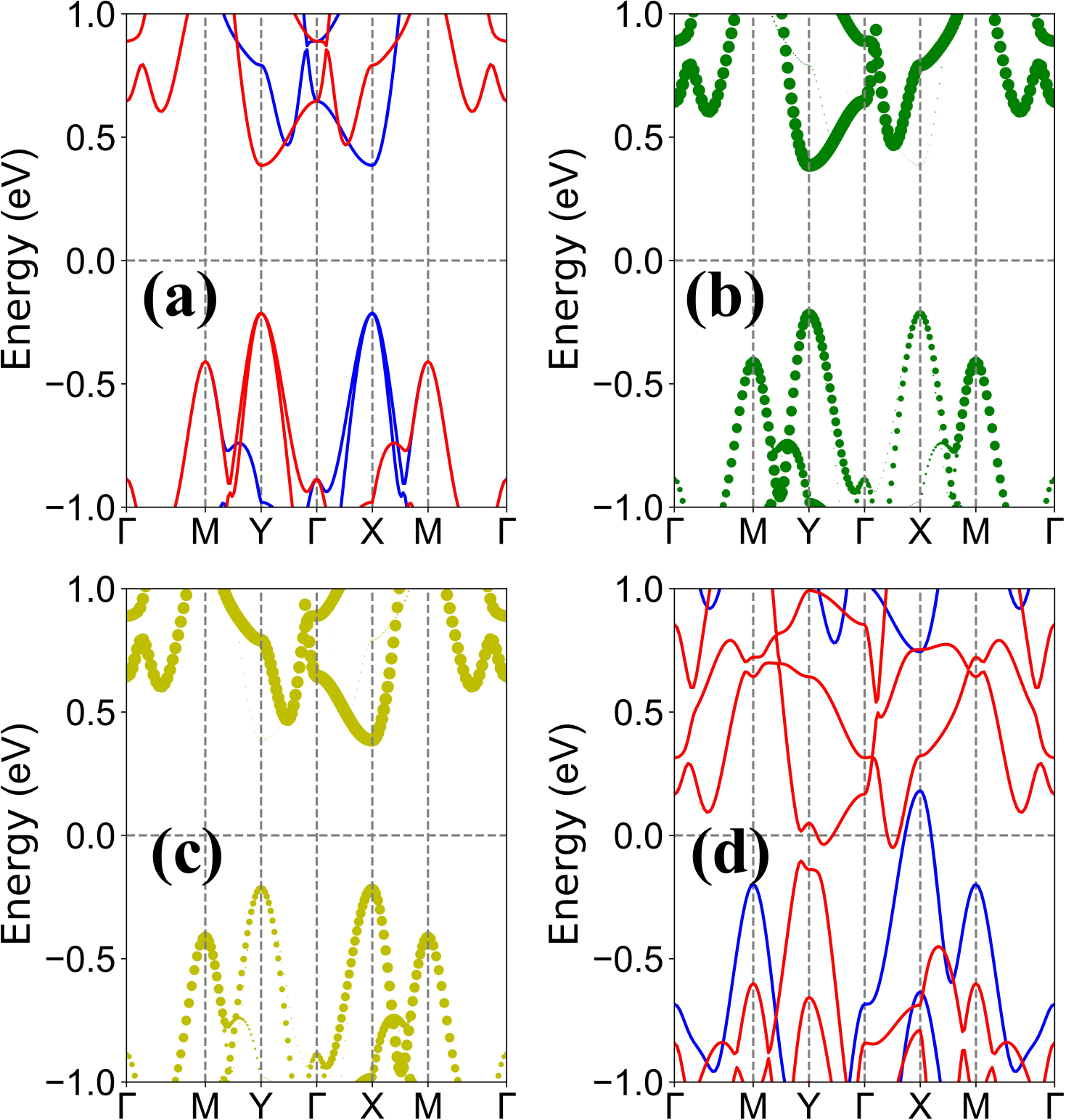}
    \caption{(Color online)For  $\mathrm{Mg(CoN)_2}$, (a):  the spin-polarized band structures without electric field; (b, c): the upper- and lower-layer Co projected band structures without electric field; (d): the spin-polarized band structures at $E$=+0.25  $\mathrm{V/{\AA}}$.  In (a, d), the blue and red curves denote the spin-up and spin-down characteristic bands, respectively, while the purple indicates the spin-degenerate bands. In (b, c), the size of the circles is proportional to the atomic weight.}\label{d}
\end{figure}

\textcolor[rgb]{0.00,0.00,1.00}{\textbf{Computational detail.---}}
The spin-polarized first-principles calculations  are carried out within density functional theory (DFT)\cite{1,1-11} and  the projector augmented-wave (PAW) method by using the Vienna Ab Initio Simulation Package (VASP)\cite{pv1,pv2,pv3}.  We use the  Perdew-Burke-Ernzerhof generalized gradient approximation (GGA)\cite{pbe}  as the exchange-correlation functional. We add Hubbard correction $U$=2 eV\cite{jp}for $d$-orbitals of Cr atoms within the
rotationally invariant approach proposed by Dudarev et al\cite{du}.  The  kinetic energy cutoff  of 550 eV,  total energy  convergence criterion of  $10^{-8}$ eV, and  force convergence criterion of 0.001 $\mathrm{eV{\AA}^{-1}}$ are adopted to obtain accurate results.  A vacuum layer exceeding 16 $\mathrm{{\AA}}$ along the $z$-direction is employed to eliminate spurious interactions between periodic images. The BZ is sampled with a 14$\times$14$\times$1 Monkhorst-Pack $k$-point meshes for both structural relaxation and electronic structure calculations. The crystal structure and atomic-position optimizations are also carried out using GGA+$U$. In the calculations under an applied electric field, the atomic positions are optimized as well.
\begin{figure*}[t]
    \centering
    \includegraphics[width=0.92\textwidth]{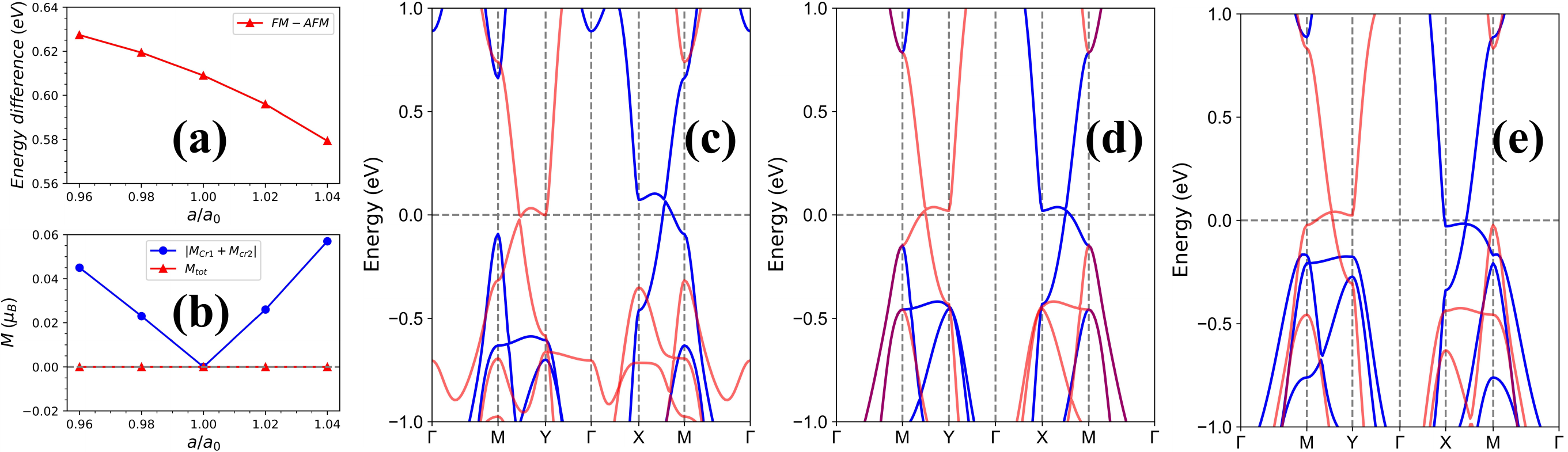}
    \caption{(Color online)(Color online)For $\mathrm{Cr_2O}$, (a): the energy difference between FM and AFM orderings  as a function of uniaxial strain $a/a_0$; (b): the absolute value of the sum of the magnetic moments of the upper- and lower-layer Cr atoms ($|M_{Cr1}+M_{Cr2}|$), along with the total magnetic moment ($M_{tot}$), as a function of  uniaxial strain $a/a_0$; (c, d, e): the spin-polarized band structures at  representative $a/a_0$=0.96,  1.00 and 1.04.  In (c, d ,e), the blue and red curves denote the spin-up and spin-down characteristic bands, respectively, while the purple indicates the spin-degenerate bands.}\label{e}
\end{figure*}

\textcolor[rgb]{0.00,0.00,1.00}{\textbf{Material realization.---}}
Monolayer $\mathrm{Cr_2O}$ has been theoretically predicted to be an AM metal with excellent stability, and it can be exfoliated from bulk CrO along the [001] direction\cite{jp}. Its crystal structures are plotted in \autoref{b} (a, b), which is composed of three monatomic planes in the sequence Cr-O-Cr and crystallizes in the  $P\bar{4}m2$ space group (No.115).  The A-type AFM state of   $\mathrm{Cr_2O}$ is the most stable magnetic
configuration as  the AM magnetic ordering.  The  optimized  equilibrium lattice constants are  $a$=$b$=3.91 $\mathrm{{\AA}}$ within GGA+$U$.

The spin-polarized band structure of $\mathrm{Cr_2O}$, along with the projected bands for the upper- and lower-layer Cr atoms, are shown in \autoref{b} (c, d, e).
It is clearly seen that $\mathrm{Cr_2O}$ is an AM semimetal, exhibiting $d$-wave symmetric spin splitting due to [$C_2$$\parallel$$S_{4z}$] symmetry.
These provide the essential basis for transforming $\mathrm{Cr_2O}$ into an FC-FIM metal via electric field or uniaxial strain tuning.
According to \autoref{b} (d, e), the bands of  two high-symmetry paths connected by $C_{4z}$ symmetry exhibit clearly distinct weights of the upper and lower Cr atoms, offering favorable conditions for band-structure tuning via an electric field.

First, we consider the modulation of $\mathrm{Cr_2O}$ by an electric field. An electric field can generate a layer-dependent electrostatic potential, rendering the upper and lower Cr atoms inequivalent and further breaking the $S_{4z}$ symmetry\cite{zg2,yz}. Viewed in reciprocal space, the field shifts the bands that carry distinct upper- and lower-layer Cr weights relative to each other, inducing new electronic states.  The energy difference between FM and AFM orderings  as a function of electric field $E$ are plotted in \autoref{c} (a).  Within the range of electric field considered, the AFM ordering remains the ground state, ensuring that $\mathrm{Cr_2O}$ retains its altermagnetism. The absolute value  ($|M_{Cr1}+M_{Cr2}|$) of the sum of the magnetic moments of the upper- and lower-layer Cr atoms  along with the total magnetic moment as a function of electric field $E$ are plotted in \autoref{c} (b). It is clear that the total magnetic moment of $\mathrm{Cr_2O}$ remains zero within considered $E$ range, satisfying the zero-net-moment requirement of a fully-compensated ferrimagnet\cite{f4}.  As the $E$ is applied, the $|M_{Cr1}+M_{Cr2}|$ deviates from zero and gradually increases, which from another perspective indicates the emergence of fully-compensated ferrimagnetism.

The spin-polarized band structures at  representative $E$=+0.00,  +0.20 and +0.40  $\mathrm{V/{\AA}}$ are plotted in \autoref{c} (c, d, e).
Under an applied electric field, $\mathrm{Cr_2O}$ remains metallic, but the bands of  two high-symmetry paths connected by $C_{4z}$ symmetry exhibit pronounced asymmetry, and the spin degeneracy along the $\Gamma$-M path is lifted. In other words, the spin-splitting symmetry of $\mathrm{Cr_2O}$ switches from $d$-wave to $s$-wave character. Upon reversal of the electric field direction, the corresponding change of the bands of  two high-symmetry paths connected by $C_{4z}$ symmetry is also reversed (see FIG.S1\cite{bc}). The $\mathrm{Cr_2O}$ exhibits zero-net magnetic moment, metallicity, and $s$-wave spin splitting under an applied electric field, enabling it to transform into an FC-FIM metal.

According to our recent proposal\cite{qq7-12}, if one layer of a bilayer system with AFM interlayer coupling is a UMS,  the system's band gap can be  closed by an electric field and realize an FC-FIM metal. This strategy can also be applied to A-type AFM AM monolayer. If the projected band structure of one magnetic-atom layer exhibits unipolar character, its gap can likewise be closed by an electric field, yielding an FC-FIM metal.
We take monolayer $\mathrm{Mg(CoN)_2}$\cite{yz} as an example to validate our proposal. Monolayer $\mathrm{Mg(CoN)_2}$ has been confirmed to be an A-type AFM AM semiconductor, and its crystal structure is shown in FIG.S2\cite{bc} with  the  $P\bar{4}m2$ space group (No.115). The spin-polarized band structure of $\mathrm{Mg(CoN)_2}$, along with the projected bands for the upper- and lower-layer Co atoms, are shown in \autoref{d} (a, b, c).
By comparing \autoref{d} (a) with  \autoref{d} (b) or  \autoref{d} (a) with  \autoref{d} (c), it can be inferred that the projected band structure of one Co layer exhibits unipolar character (Near the valence band maximum (VBM) and  conduction band minimum (CBM), the bands are dominated by the same spin character.), which provides the essential condition for realizing an FC-FIM metal through electric-field control\cite{qq7-12}. When an electric field of  $E$=+0.25  $\mathrm{V/{\AA}}$ is applied, the total magnetic moment of $\mathrm{Mg(CoN)_2}$ is zero, its band structure exhibits metallic character and spin splitting (see \autoref{d} (d)), and then it indeed turns into an FC-FIM metal.

\begin{figure}[t]
    \centering
    \includegraphics[width=0.46\textwidth]{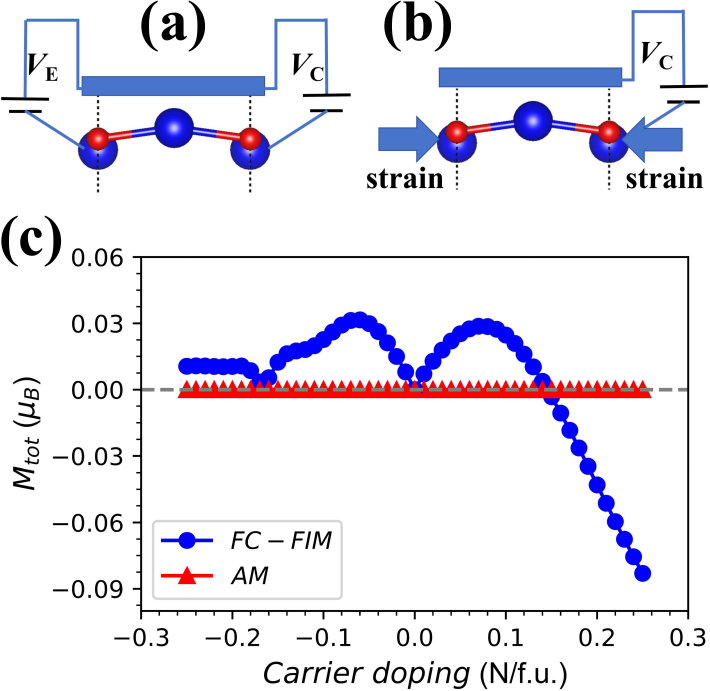}
    \caption{(Color online)For $\mathrm{Cr_2O}$, (a): by using dual-gate technology, the electric field $E$ with $V_E$ and the charge-carrier density $N$ with $V_C$ can be tuned independently,  and  the  $\mathrm{Cr_2O}$ can be turned into a fully-compensated ferrimagnet, exhibiting a net magnetic moment;  (b): The $V_E$ in (a) can be replaced by uniaxial strain to achieve fully-compensated ferrimagnetism, and  generate a net magnetic moment; (c): the total magnetic moment as a function of carrier concentration for the unstrained case (AM) and for $a/a_0$=1.04 (FC-FIM). }\label{f}
\end{figure}

Next, we discuss the effect of uniaxial strain on the electronic structure of $\mathrm{Cr_2O}$. We use $a/a_0$ to simulate uniaxial strain along the $x$-direction, and  the $a/a_0$$<$1 ($a/a_0$$>$1) means the compressive (tensile) strain, where   $a$ and $a_0$ are the strained and  unstrained lattice constants.
Uniaxial strain directly alters the crystal structure, thereby breaking the $S_{4z}$ symmetry of $\mathrm{Cr_2O}$. Analogous strain engineering has been applied to other AM semiconductors to induce valley polarization\cite{k6,k7,k7-1,k7-2,k7-3}.
The $a/a_0$ dependence of the energy difference between FM and AFM orderings is displayed in \autoref{e}(a). Throughout the examined range, the AFM phase stays lower in energy, so $\mathrm{Cr_2O}$ preserves AM ground state. \autoref{e}(b) shows the absolute value  ($|M_{Cr1}+M_{Cr2}|$) of the sum of the magnetic moments on the upper- and lower-layer Co atoms, together with the total magnetic moment, as a function of $a/a_0$. The total magnetic moment remains exactly zero, satisfying the fully-compensated ferrimagnetic requirement, whereas $|M_{Cr1}+M_{Cr2}|$ progressively departs from zero under increasing  both compressive and tensile strain, providing direct evidence for the strain-induced emergence of fully-compensated ferrimagnetism\cite{f4}.

The spin-polarized band structures at  representative $a/a_0$=0.96,  0.00 and 1.04  are shown in \autoref{e} (c, d, e).
Upon applying strain, $\mathrm{Cr_2O}$  retains its metallic character; however, the bands along the two high-symmetry paths related by $C_{4z}$ symmetry show a marked asymmetry, and the spin degeneracy along the $\Gamma$-M path is removed. Consequently, the spin-splitting symmetry of $\mathrm{Cr_2O}$  transitions from a $d$-wave to an $s$-wave character. It is also found that the band modifications induced by tensile-to-compressive strain mirror those produced by reversing the electric field direction. If uniaxial strain is applied along the $y$-direction, similar results are obtained, and the sole difference is  that the bands on the two paths linked by $C_{4z}$ symmetry are interchanged.
Subject to strain,  $\mathrm{Cr_2O}$   retains metallicity and zero-net magnetic moment, while developing $s$-wave spin splitting, enabling its transition into an FC-FIM metal.

Finally, we examine how charge-carrier doping affects the total magnetic moment of $\mathrm{Cr_2O}$.
In 2D AM semiconductor, the emergence of a net magnetic moment under the cooperative action of uniaxial strain and charge-carrier doping is designated as piezomagnetism\cite{k6}. Here,  we emphasize that the underlying origin of this effect is the strain-induced transition from an AM to an FC-FIM  state under uniaxial deformation (According to \autoref{a} (e, f), only fully-compensated ferrimagnets can produce a net magnetic moment upon charge-carrier doping.). Under biaxial strain,  the system remains AM phase  and no piezomagnetic response emerges.
Analogously, we can also introduce an electromagnetic counterpart: by driving an altermagnet (more generally, including  $PT$-antiferromagnet) into a fully-compensated ferrimagnet with an electric field, subsequent charge-carrier doping can produce  a net magnetic moment. This electrically controlled magnetoelectric phenomenon is designated electromagnetism.

Dual-gating technology has been implemented in graphene to enable tunable control of proton transport and
hydrogenation\cite{dg}.  Dual-gating by applying electrostatic gate potentials to both surfaces of a 2D system enables independent control of the electric field $E$ and the charge-carrier density $N$ (see \autoref{f} (a)), and the electromagnetism can be realized in  $\mathrm{Cr_2O}$.  The piezomagnetism in $\mathrm{Cr_2O}$ can also be achieved through the combined application of uniaxial strain and single-gate electrostatic control (see \autoref{f} (b)).
Owing to computational constraints, the simultaneous imposition of an external electric field and charge-carrier doping cannot be implemented within the VASP code.
Herein, we exclusively impose uniaxial strain together with charge-carrier doping to verify the emergence of a net magnetic moment in $\mathrm{Cr_2O}$.

The total magnetic moment as a function of carrier concentration for the unstrained case and for $a/a_0$=1.04 are plotted in \autoref{f} (c).
It is evident that, in the absence of strain,  $\mathrm{Cr_2O}$ remains an altermagnet exhibiting no net magnetic moment, whereas under applied strain it transitions to a fully-compensated ferrimagnet, thereby generating a finite net magnetic  moment. Across the entire doping window investigated under finite strain, the AFM ordering of  $\mathrm{Cr_2O}$ persists as the ground state, retaining its FC-FIM character (see FIG.S3\cite{bc}). To unambiguously demonstrate that a fully-compensated ferrimagnet can host a net magnetic moment under charge-carrier doping, we select the prototypical FC-FIM  semiconductor $\mathrm{CrMoC_2S_6}$\cite{v1,v2}  and $\mathrm{Cr_2CHCl}$\cite{v3,v4} for further validation of this effect. Their band structures, the energy difference between FM and AFM states  as  a function of doping concentration, and the total magnetic moment as a functions of doping concentration are presented in FIG.S4, FIG.S5 and FIG.S6\cite{bc}.
It is evident that charge-carrier doping can induce a net magnetic moment in  FC-FIM  semiconductor $\mathrm{CrMoC_2S_6}$  and $\mathrm{Cr_2CHCl}$.

\textcolor[rgb]{0.00,0.00,1.00}{\textbf{Discussion and conclusion.---}}
A key band-structure distinction between altermagnet and fully-compensated ferrimagnet is that in the former every band at the high-symmetry $\Gamma$ point remains spin-degenerate, whereas in the latter this need not be the case. Moreover, gate-controlled charge-carrier doping leaves an altermagnet with zero-net-moment, whereas the same procedure generates a finite magnetization in a fully-compensated ferrimagnet. For practical use, an FC-FIM semiconductor must be charge-carrier-doped, which will generate a net magnetic moment. In contrast, FC-FIM half-metal or  metal already possesses intrinsic carriers and can therefore retain zero-net-moment. Finally, we point out that either thermal excitation or light-induced carrier injection can also give rise to a net magnetization in a fully-compensated ferrimagnet.

In summary, we demonstrate that breaking the $C/M$ symmetry which couples spin-opposite magnetic atoms in an AM metal provides a direct route to realizing an FC-FIM metal.  First-principles calculations show that AM  $\mathrm{Cr_2O}$ can form an FC-FIM metal via electric field or uniaxial strain. By analogy with piezomagnetism, we introduce the concept of electromagnetism. We further point out that the net magnetic moment induced by combined strain or electric field with carrier doping essentially arises from the external-field-driven transition of the system from the AM state to the FC-FIM state. Our work provides a feasible strategy for  realizing FC-FIM metal, along with the connection and distinction between the AM and FC-FIM states.

\begin{acknowledgments}
This work is supported by Natural Science Basis Research Plan in Shaanxi Province of China  (2025JC-YBMS-008) We are grateful to Shanxi Supercomputing Center of China, and the calculations were performed on TianHe-2.
\end{acknowledgments}

\end{document}